\documentclass[useAMS]{mn2e}
\usepackage{amssymb,amsmath,psfig,times}
\usepackage{graphicx}
\def\gsim{ \lower .75ex \hbox{$\sim$} \llap{\raise .27ex \hbox{$>$}} }
\def\lsim{ \lower .75ex\hbox{$\sim$} \llap{\raise .27ex \hbox{$<$}} }

\def\beq{\begin{equation}}
\def\eeq{\end{equation}}

\def\fe{{\it Fermi}}

\title[Time-resolved Spectral Study Of Fermi GRBs]
{Time-resolved Spectral Study Of \fe\ GRBs Having  Single Pulses}
\author[R. Basak et al.]
{Rupal Basak\thanks{E--mail:rupalb@tifr.res.in}, 
A. R. Rao\thanks{E--mail:arrao@tifr.res.in}\\
TIFR-- Tata Institute of Fundamental Research, Mumbai 400005, India
}
\begin{document}

\date{}


\maketitle

\label{firstpage}

\begin{abstract}
We analyze gamma-ray bursts (GRBs) detected by \textit{Fermi}/Gamma-Ray Burst Monitor (GBM) and having single pulse. We fit the light
curves with a model having exponential rise and decay parts. We perform a detailed time-resolved spectroscopy using four models: 
Band, blackbody 
with a power-law (BBPL), multicolour blackbody with a power-law (mBBPL) and two blackbodies with a power-law (2BBPL). 
We find that models other than the BBPL give better $\chi_{red}^2$ for the ``hard-to-soft'' (HTS) pulses, 
while  for the ``intensity tracking'' (IT) pulses, the BBPL model is statistically as good as the other models. Interestingly, 
the energy at the peak of the spectrum resulting from the BBPL model ($\sim3kT$), is always lower than that of the $\nu F_{\nu}$  
spectrum of the Band function. The values of the low energy photon index ($\alpha$) of the Band function are 
often higher than the fundamental single particle synchrotron limit,
especially for the HTS pulses. Specifically we find two extreme cases --- 
for GRB~110817A (HTS GRB) $\alpha$ is always higher, while for GRB~100528A (IT GRB) $\alpha$ 
is always within the synchrotron regime. The PL component of the BBPL model always starts with a delay compared 
to the BB component, and it lingers at the later part of the  
prompt emission phase. For three HTS GRBs, namely, GRB~081224, GRB~100707A and GRB~110721A 
this behaviour is particularly significant and interestingly there are reported LAT detections for them. 
Finally, we argue that various evidences hint that 
neither BBPL nor Band model is acceptable, while 2BBPL and mBBPL are the most 
acceptable models for the set of GRBs we have analyzed.

\end{abstract}
\begin{keywords}
 radiation mechanisms: non-thermal -- radiation mechanisms: thermal -- methods: data analysis -- methods: observational -- methods: statistical -- gamma-ray burst: general.
\end{keywords}

\section{Introduction}
A gamma-ray Burst (GRB) is the most luminous event in the universe. In a few seconds it produces an enormous amount of electromagnetic
energy, which is comparable to the integrated emission of the Sun over ten billion years (Meszaros 2006). 
Most of this energy is released in the first few seconds in
the form of gamma rays (keV to MeV), known as the prompt emission phase. In some GRBs very high energy 
(GeV) emission is also observed during the prompt phase. Later the afterglow of a GRB can be observed at all electromagnetic 
frequencies from the radio wavelengths up to x-rays and even gamma rays. Numerous satellite experiments have been performed to pin down the 
mechanism of GRBs. For ten long years (1991-2000), a great wealth of prompt emission data was collected by the 
Burst and Transient Source Experiment (BATSE; Fishman et al. 1994) onboard the Compton Gamma-Ray Observatory (\textit{CGRO}). With 
\textit{Swift} (Barthelmy et al. 2005; Gehrels et al. 2004) and \textit{Fermi} (Meegan et al. 2009), launched respectively in 2004 and 2008, 
we have entered into the modern era of GRB research. \textit{Swift}, with its fantastic slewing capability, an unprecedented 
localization accuracy of the primary GRB detector (Burst Alert Telescope -- BAT, a large field of view gamma ray monitor) 
and the high spatial resolution instruments for the afterglow studies has enabled the measurement of the redshift for 
many GRBs. \textit{Fermi}, on the other hand, has enabled a detailed spectral study during the prompt emission phase, and
discovered the very high energy (GeV) emission both during the prompt and afterglow phase for many bursts (very few such
events were known before \textit{Fermi}. See e.g., Gonz{\'a}lez et al. 2003).
Both these satellites have supplied extremely valuable data for the identification of the 
working mechanism of GRBs. But, even after nearly fifty years of discovery, various issues remain puzzling, most notably the 
prompt emission mechanism. 

In the preliminary model of a GRB, known as the ``standard fireball model'' (Goodman 1986; Paczynski 1986),
{the radiation is supposed to come from a photosphere.} 
The instantaneous spectrum is predicted to be a blackbody (BB) and the light curve (LC) is a simple pulse. Though we do see a GRB having  
a single pulse, there is a variety of LC profiles. There are also differences in the predicted and observed spectrum.
While the BB spectral shape in the photon space is a hump 
with a narrow peak, which can be approximated with a low energy power-law with a +1 index and a steep power-law with a negative index 
at high energy, GRB prompt emission spectra usually have non-thermal spectral shape characterized with negative power-law 
indices ($\sim-1$). Though there is a provision for geometric broadening (Goodman 1986), 
the spectrum is still far from what is observed. To overcome this difficulty, it is assumed that the radiation is not coming
from the photosphere, but it is produced in the internal shocks (IS) of the ejecta via synchrotron radiation (Rees \& Meszaros 1994;
Woods \& Loeb 1995; Sari \& Piran 1997; Kobayashi et al. 1997). This model is phenomenologically represented by the featureless Band
function (Band et al. 1993), which is a smoothly joined power-laws with two indices --- $\alpha$ as the lower energy 
index and $\beta$ as the higher energy index. In the EF(E) representation, the function peaks at an energy ($E_{\rm peak}$),
provided that $\beta<-2$, which can be violated for very hard spectra.
The prompt emission data, whether time-integrated or time-resolved, is adequately fit by the Band function (Kaneko et al. 2006; 
Nava et al. 2011; Zhang et al. 2011). However, the Band function is a phenomenological function and the actual spectrum may have
rich structure. In many studies, additional components to the Band function (Preece et al. 1996; 
Meszaros \& Rees 2000; Gonz{\'a}lez et al. 2003; Shirasaki et al. 2008; Abdo et al. 2009, Ackermann 
et al. 2010; Ackermann et al. 2011; Guiriec et al. 2011; Axelsson et al. 2012; Guiriec et al. 2013) and 
even alternative models (Ghirlanda et al. 2003; Ryde 2004; Ryde \& Pe'er 2009; Pe'er \& Ryde 2011) have 
been suggested. The reason for resorting to alternative description is that the internal shock model demands that in the electron
slow cooling regime, the photon index ($\alpha$) should be less than -2/3 (Preece et al. 1998), the single particle synchrotron limit. 
GRBs are usually expected to be in the fast cooling regime where the photon index should be near -3/2 (Cohen et al. 1997). 
These two limits are referred to as the ``synchrotron lines of death''. The value of $\alpha$ is often found to be higher 
than the -2/3 (Crider et al. 1999). Moreover, the IS model has the following issues. The efficiency of converting the 
kinetic energy to the internal energy and then radiating is rather low (less than 20\%; Piran 1999; 
but, see also Nemmen et al. 2012). The situation of an inner shell moving faster than an outer shell, 
which is a necessary condition for the IS to produce, is unstable (Waxman \& Piran 1994).

These difficulties of the IS model have instigated the search  for alternate models. 
Some of them involve looking back into the prediction 
of the original fireball model. For example, Ryde (2004) has shown for a few BATSE GRBs, having single pulse, that the 
instantaneous spectrum is consistent with either a BB or a BB with a power-law (BBPL) spectrum (see also Ryde \& Pe'er 2009).
The reasons to choose single pulse are two-fold --- (a) the original 
fireball model predicts single pulse and instantaneous BB radiation; (b) if the spectral evolution is a pulse property rather 
than a burst property, a single pulse is an ideal system to study the spectral evolution and then one can apply the knowledge 
of a single pulse to a more complex GRB. The BBPL model has sometimes shown superiority over the Band function 
in the BATSE data. This is a very promising result because, this is the first step towards identifying the physical mechanism of 
the radiation. However, later it was found that more complex models may be required to fit a wider spectrum provided by 
Fermi (Ryde et al. 2010; Burgess et al. 2011; Guiriec et al. 2013; Basak \& Rao 2013 --- BR13, hereafter). It is shown that a model
consisting of a Band and a blackbody model (Band+BB) is required to fit some of the long (Guiriec et al. 2011; Axelsson et al. 2012) 
and short (Guiriec et al. 2013) GRBs. Interestingly, in a Band+BB fit, 
the parameters of the Band function are compatible with the 
synchrotron model prediction, while they are not for a fit with a Band only function.
Band+BB scenario with a subdominant BB is not compatible with the genuine fireball model and 
requires a highly magnetized outflow  close to the source and a low magnetization at large radii to explain the observations. In addition, 
the magnetization seems to vary from burst to burst (Guiriec et al. 2013). In this context, 
it is important to examine the Fermi GRBs having single pulses employing various models. Lu et al. (2012; Lu12 hereafter) 
have analyzed a set of Fermi GRBs. They have paid special attention to the single pulses. Their motivation, however, was 
to find the variation of $E_{\rm peak}$ of Band function. In this study, we examine various models and their parameter evolution 
for GRBs having single pulses.

The organization of this paper is as follows. In Section 2, we describe our sample, followed by data analysis techniques.
In Section 3, we show our results. The major findings are summarized and discussed in Section 4.
\begin{table*}
\begin{scriptsize}

\caption{Parameters for Norris model fit to the light curve of the GRB pulses. We also report the time interval ($t_1$ to $t_2$)
in which we have performed our analysis, number of time bins (n) within the time interval and the best detectors used for 
time-resolved spectroscopy.  }

\begin{tabular}{c|ccccccc|ccc}
\hline 

GRB & \multicolumn{7}{c|}{Norris Model parameters} & \multicolumn{3}{c}{Specification of time-resolved analysis}\\

\cline{2-11}

    & $t_s$ (s) & $\tau_1$ (s) & $\tau_2$ (s) & $\chi^2_{red} (dof)$ & p (s) & w (s) & $\kappa$ & $t_1$,$t_2$ (s) & n & Detectors used \\
\hline
\hline 
081224 & $-1.53_{-0.26}^{+0.34}$ & $3.60_{-1.07}^{+0.99}$ & $3.30_{-0.21}^{+0.27}$ & 6.80 (13) & $1.90\pm 0.10$ & $7.5\pm 0.7$& $0.44\pm 0.03$& -0.5, 19.6 & 15 & n6, n7, n9, b1\\
090809B & $-1.74_{-0.38}^{+0.31}$ & $9.35_{-1.85}^{+2.59}$ & $2.50_{-0.17}^{+0.17}$ & 5.88 (13) & $3.10\pm 0.09$& $7.4\pm 0.6$& $0.34\pm 0.02$& -4.0, 19.3 & 15 & n3, n4, n5, b0\\
100612A & $-8.43_{-2.65}^{+1.61}$ & $156.4_{-57.4}^{+137.6}$ & $0.88_{-0.17}^{+0.14}$ & 16.8 (9) & $3.30\pm 0.06$ & $6.5\pm 1.5$& $0.14\pm 0.02$& -2.0, 17.2 & 11 & n3, n4, n8, b0\\
100707A & $-2.5_{-2.6}^{+3.2}\times 10^{-2}$ & $0.37_{-0.05}^{+0.04}$ & $7.19_{-0.16}^{+0.18}$ & 4.01 (233) & $1.60\pm0.07$ & $9.9\pm0.28$& $0.72\pm0.01$&  -1.0, 22.6 & 18 & n4, n7, n8, b1 \\
110721A & $-0.62_{-0.15}^{+0.08}$ & $1.58_{-0.46}^{+0.67}$ & $1.59_{-0.23}^{+0.15}$ & 1.32 (229) & $0.96\pm0.10$ & $3.55\pm0.62$& $0.45\pm0.03$&  -1.0, 10.6 & 15 & n6, n7, n9, b1 \\
        & $1.75_{-0.07}^{+0.10}$ & $0.079_{-0.053}^{+0.074}$ & $6.75_{-0.66}^{+0.86}$ &  & $2.48\pm0.12$ & $8.08\pm1.04$& $0.83\pm0.05$&  &  &  \\
110817A & $-0.62_{-0.42}^{+0.30}$ & $1.11_{-0.66}^{+1.31}$ & $1.76_{-0.24}^{+0.24}$ & 1.85 (8) & $0.78\pm 0.12$ & $3.6\pm 0.8$& $0.49\pm 0.08$& -0.3, 7.1 & 8 & n6, n7, n9, b1\\
081207 & $-8.55_{-2.21}^{+3.30}$ & $14.14_{-6.10}^{+9.10}$ & $50.79_{-10.22}^{+11.10}$ & 1.36 (114) & $18.25\pm 3.36$ & $90\pm 21$& $0.57\pm 0.06$& -7.0, 76.0 & 20 & n1, n9, na, b1\\
       & $-9.64$ & $304.7$ & $9.24_{-2.20}^{+3.50}$ & & $43.42 \pm 1.19$ & $45$& $0.20$&  & &\\
090922A & $-0.39_{-0.19}^{+0.15}$ & $0.62_{-0.27}^{+0.38}$ & $4.00_{-0.30}^{+0.30}$ & 6.23 (15) & $1.18\pm 0.18$ & $6.4\pm 0.7$& $0.62\pm 0.05$& -4.0, 13.4 & 8 & n0, n6, n9, b1\\
100528A & $-32.6_{-2.36}^{+2.36}$ & $1262.0_{-61.1}^{+60.9}$ & $1.29_{-0.38}^{+0.14}$ & 6.56 (21) & $7.70\pm 0.06$ & $14.5\pm 3.0$& $0.09\pm 0.004$& -3.0, 21.4 & 16 & n6, n7, n9, b1\\

\hline
\end{tabular}
\label{t1}
\end{scriptsize}

\end{table*}

\section{Data Selection and Analysis}
We select our sample from Lu12. They reported 51 long and 11 short bursts in \textit{Fermi}/GBM catalog
till 2011 August 31. This sample was selected by requiring the following criteria. i) GRBs for which less than 5 time-resolved
bins could be obtained for a signal-to-noise ratio of 35 are neglected. ii) A lower limit on the fluence, (calculated 
in the GBM energy range --- 8 to 900 keV), is put to select only bright GRBs. This limit is 10$^{-5}$ erg cm$^{-2}$ for 
long GRBs and 8$\times$10$^{-7}$ erg cm$^{-2}$ for short GRBs. Lu12 have reported 8 long GRBs which have single pulses. 
We use these GRBs for our analysis. Lu12 further divided the sub-sample into two categories, namely, ``hard-to-soft'' (HTS; GRB~081224,
GRB~090809B, GRB~100612A, GRB~100707A and GRB~110817A) and ``intensity tracking'' (IT; GRB~081207, GRB~090922A and GRB~100528A). 
This classification is done depending on the time evolution of $E_{\rm peak}$ --- ``HTS'' if the evolution is strictly
descending in time, and ``IT'' if the evolution follows the flux evolution. We have added GRB 110721A to this 
catalog. This GRB has a hard-to-soft evolution and high energy GeV detection. This GRB falls short in the fluence 
criteria of Lu12. However, we could obtain 15 time-resolved bins, because the peak flux of this GRB is high (see 
Axelsson et al. 2012).

\begin{figure*}\centering
{

\includegraphics[width=6.0in]{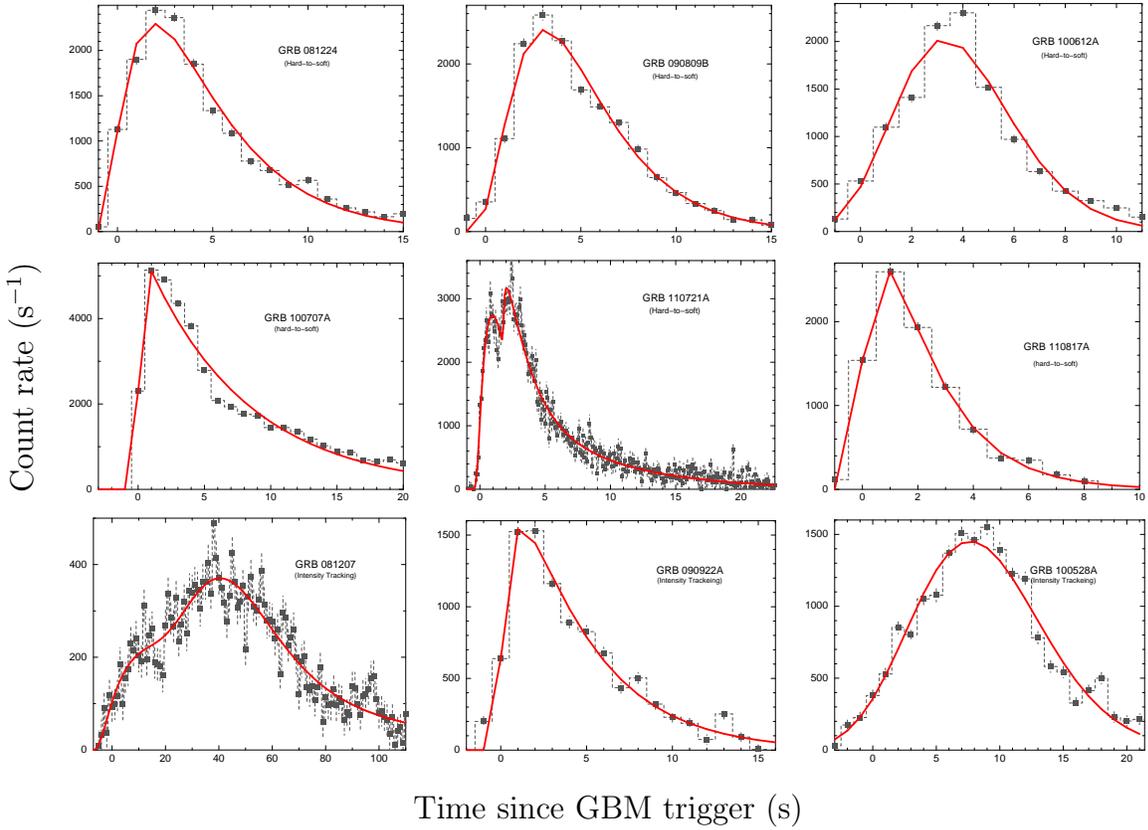} 

}
\caption{Background subtracted light curves (LCs) of the GRBs in 8-900 keV energy range for the NaI detector in which the count
rate is maximum. The upper 6 panels are ``hard-to-soft'' pulses and the lower
3 panels are ``intensity tracking'' pulses. The LCs are fitted with Norris' exponential model (Norris et al. 2005).
The corresponding values of this model are shown in Table~\ref{t1}.
}
\label{lc}
\end{figure*}

\subsection{Timing Analysis}
The light curves (LCs) of the GRBs  are generated in the full energy range (8-900 keV)
for the NaI detector in which the count
rate is maximum (Figure 1).
To fit the LCs of these GRBs, we use the exponential model (Norris et al. 2005). The exponential model is a
four parameter model and adequately fits the GRB LC (see Norris et al. 2005; Rao et al. 2011; Basak \& Rao et al. 2012a; b). 
\begin{equation}
I(t>t_{s})=A_n\lambda exp\{-\tau_{1}/(t-t_{s})-(t-t_{s})/\tau_{2}\}\label{1}
\end{equation}
Here $t_{s}$ is the pulse start time, $\tau_{1}$, $\tau_{2}$ are characteristic time scales denoting the rise and decay period of the 
pulse, and $A_n$ is the pulse amplitude. These are the four parameters of the model and  $\lambda=exp\left(2\mu\right)$, where 
$\mu=\left(\tau_{1}/\tau_{2}\right)^{\frac{1}{2}}$. Various other quantities characterizing a pulse can be derived from 
these parameters e.g., peak position (p), pulse width (w; separation between the times where intensity is 1/e of the maximum), 
and asymmetry ($\kappa$; see Norris et al. 2005 for details). We calculate the error in the model parameters at nominal 90\% 
confidence level ($\triangle\chi^{2}$=2.7) and use them to calculate the errors in the derived parameters.

\begin{figure*}\centering
{

\includegraphics[width=6.0in]{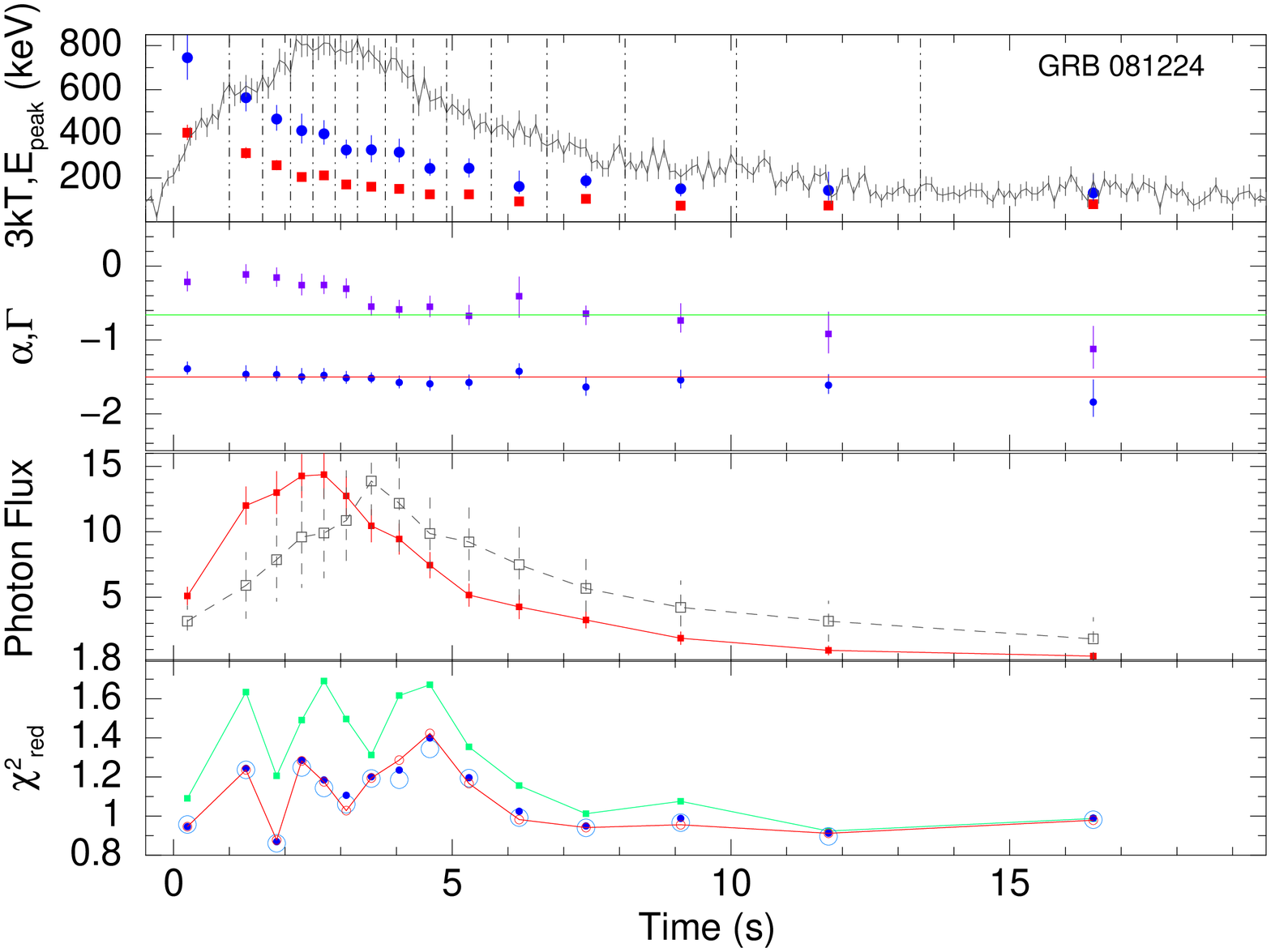} 

}
\caption{GRB 081224. From upper panel, panel 1: variation of 3kT (red filled boxes) and $E_{\rm peak}$ (blue filled circles) with time.
Shown in background is the corresponding lightcurve plotted in the same scale as in Figure~\ref{lc}. The dot-dashed
lines denote the time intervals used for spectral analysis. The parameters are shown at the mean time of each interval.
Panel 2: $\alpha$ (violate boxes) of Band function and $\Gamma$ (blue circles) of BBPL model. The lines denote 
the synchrotron ``lines of death'' (one at -3/2 and another at -2/3).
Panel 3: Flux (photons cm$^{-2}$ s$^{-1}$) variation of BB (filled boxes joined by straight lines) and PL (open boxes joined 
by dashed line). The power-law flux are scaled by the ratio of total BB flux and total power-law flux. 
Panel 4: $\chi^{\rm2}_{\rm red}$ for BBPL (green filled boxes), Band (red open circles), mBBPL (blue filled circles) and
2BBPL (largest open circles)
}
\label{grb1}
\end{figure*}

\begin{figure*}\centering
{

\includegraphics[width=6.0in]{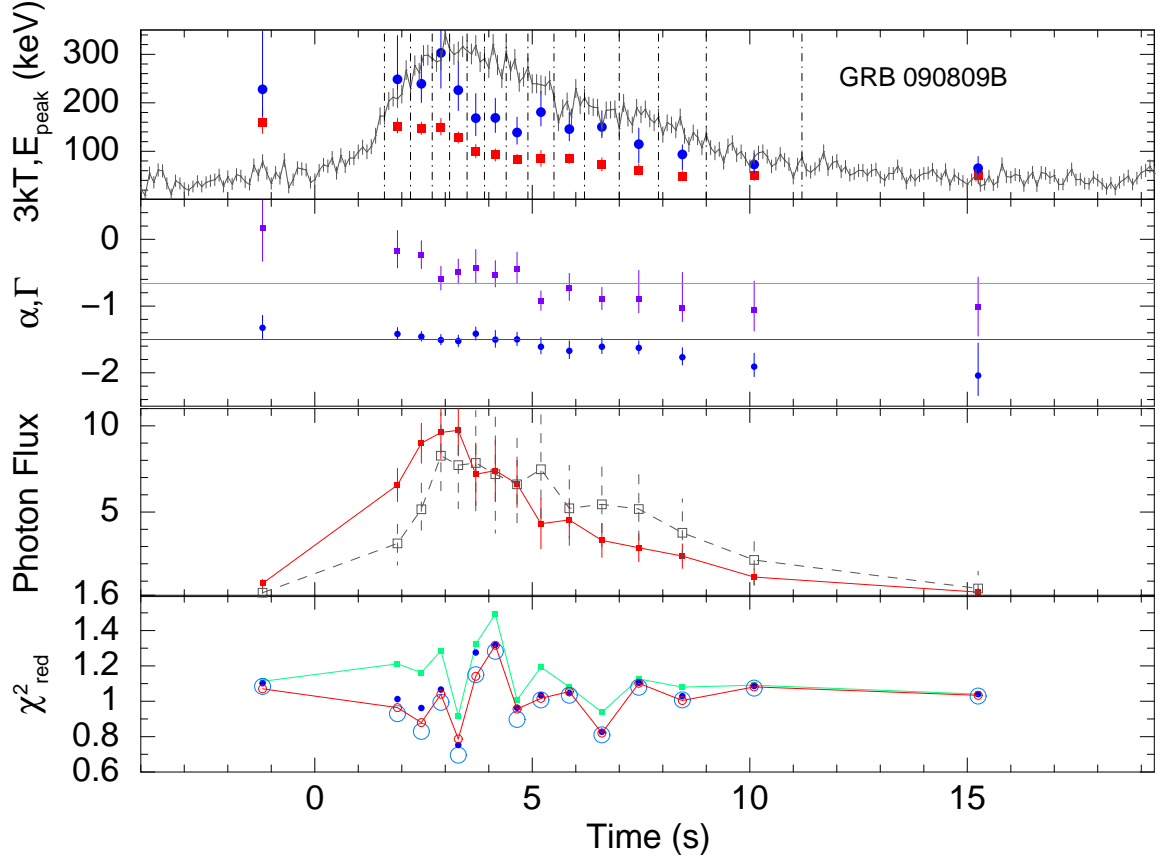} 

}
\caption{GRB 090809B. Symbols used are the same as in Figure~\ref{grb1}
}
\label{grb2}
\end{figure*}

\begin{figure*}\centering
{

\includegraphics[width=6.0in]{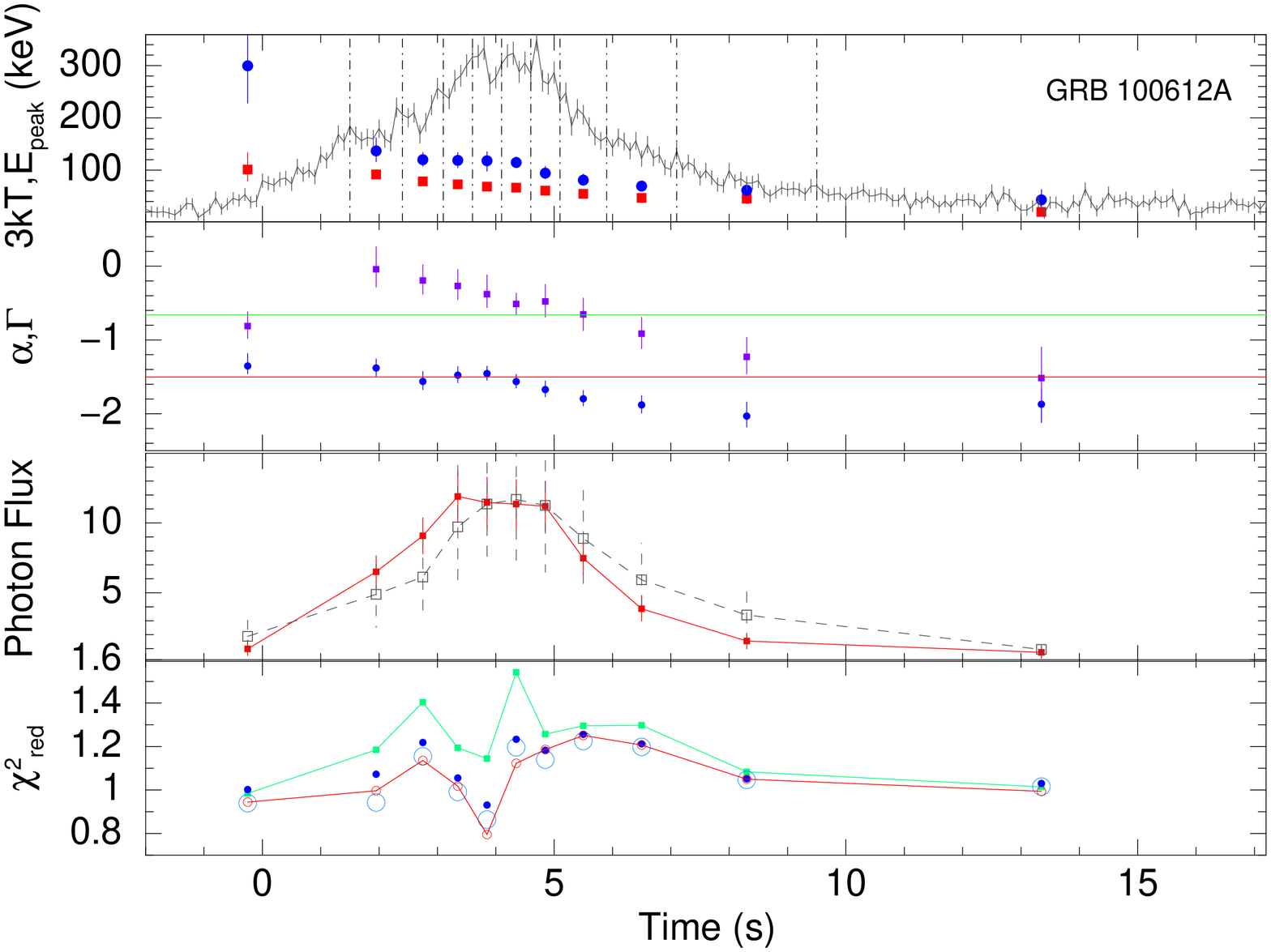} 

}
\caption{GRB 100612A. Symbols used are the same as in Figure~\ref{grb1}
}
\label{grb3}
\end{figure*}

\begin{figure*}\centering
{

\includegraphics[width=6.0in]{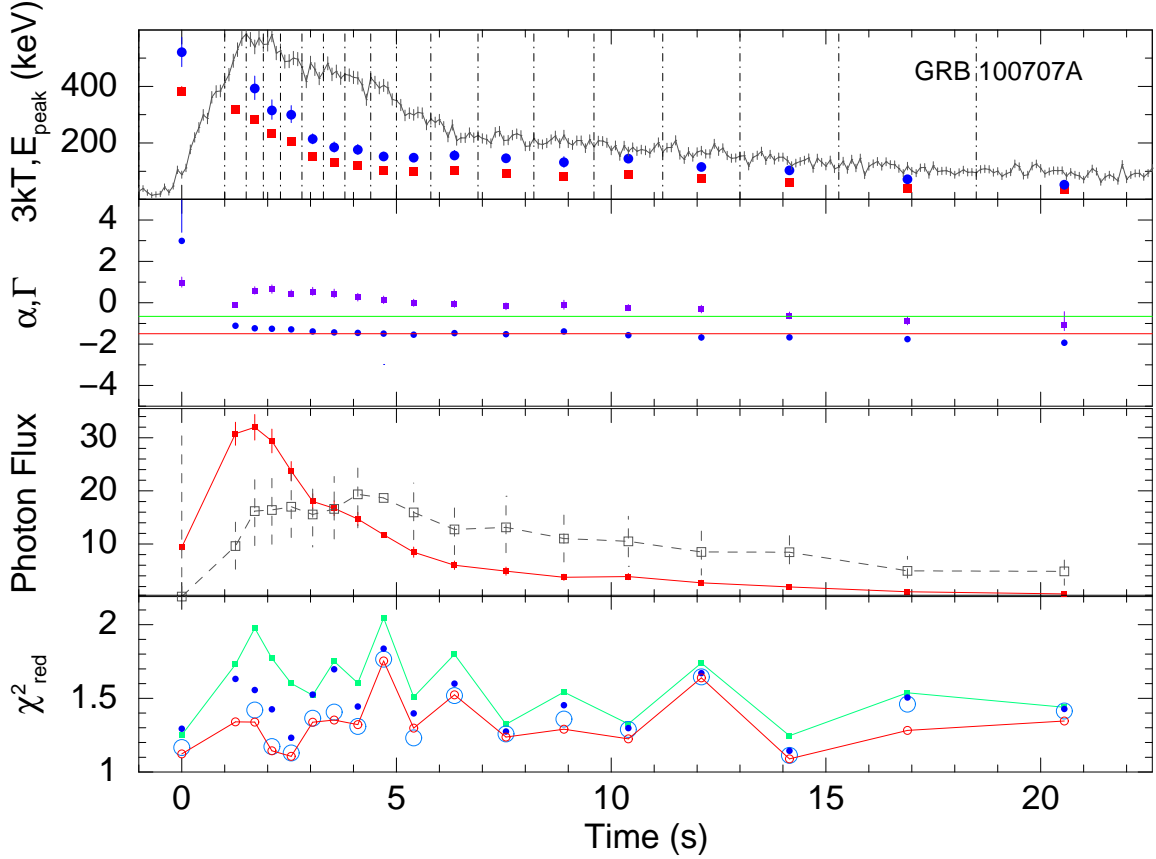} 

}
\caption{GRB 100707A. Symbols used are the same as in Figure~\ref{grb1}
}
\label{grb4}
\end{figure*}

\begin{figure*}\centering
{

\includegraphics[width=6.0in]{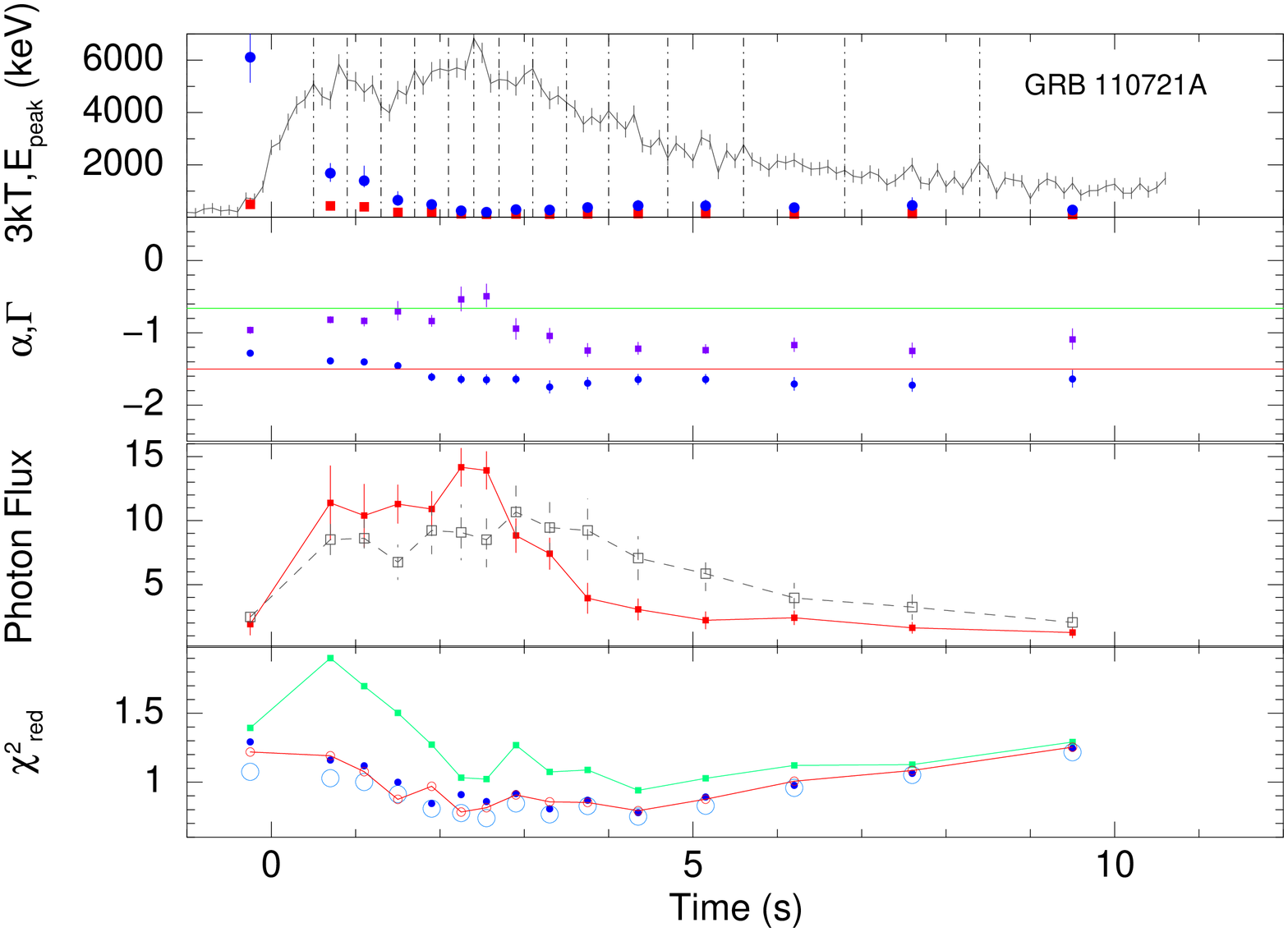} 

}
\caption{GRB 110721A. Symbols used are the same as in Figure~\ref{grb1}
}
\label{grb5}
\end{figure*}

\begin{figure*}\centering
{

\includegraphics[width=6.0in]{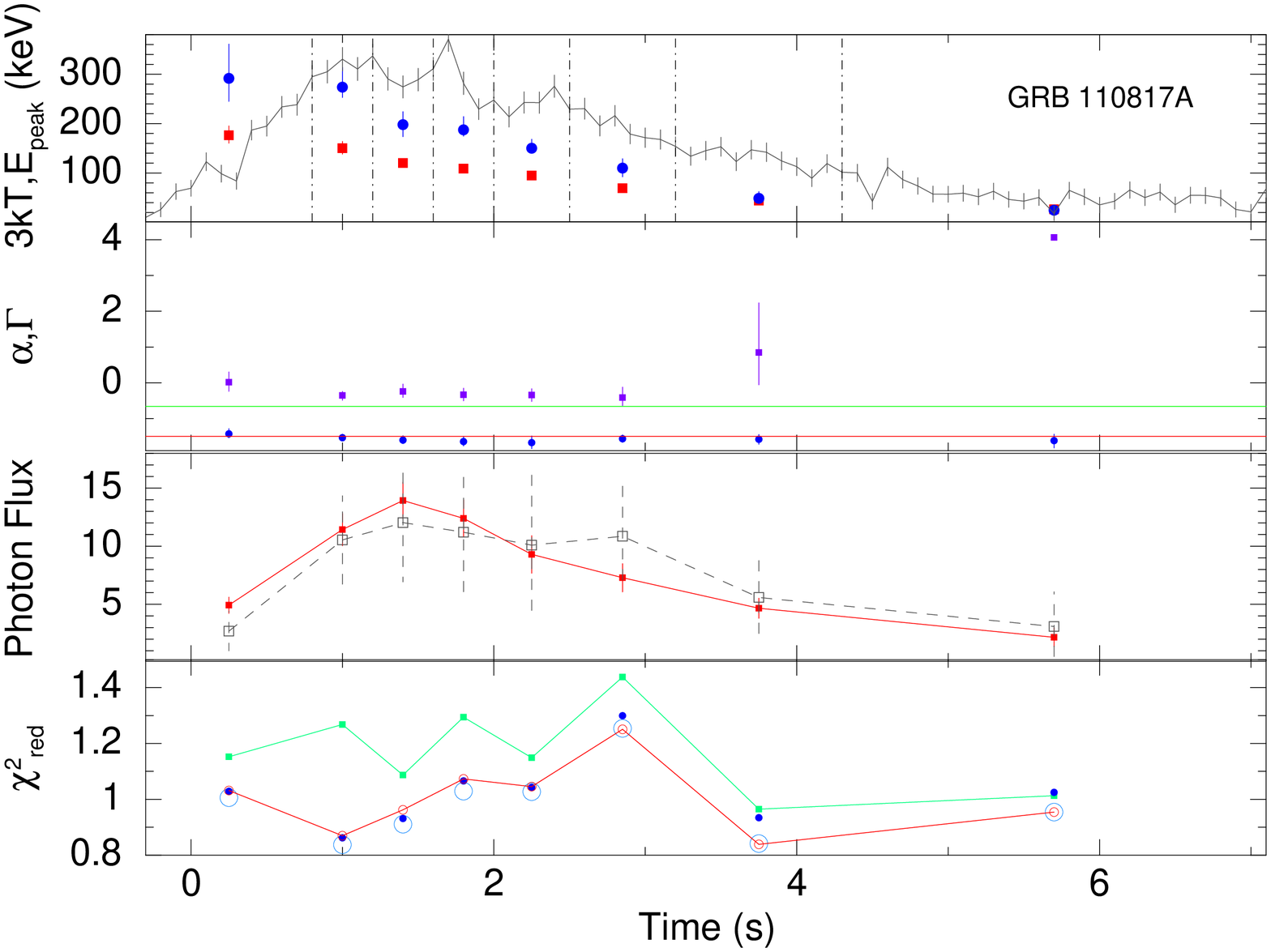} 

}
\caption{GRB 110817A. Symbols used are the same as in Figure~\ref{grb1}
}
\label{grb6}
\end{figure*}

\begin{figure*}\centering
{

\includegraphics[width=6.0in]{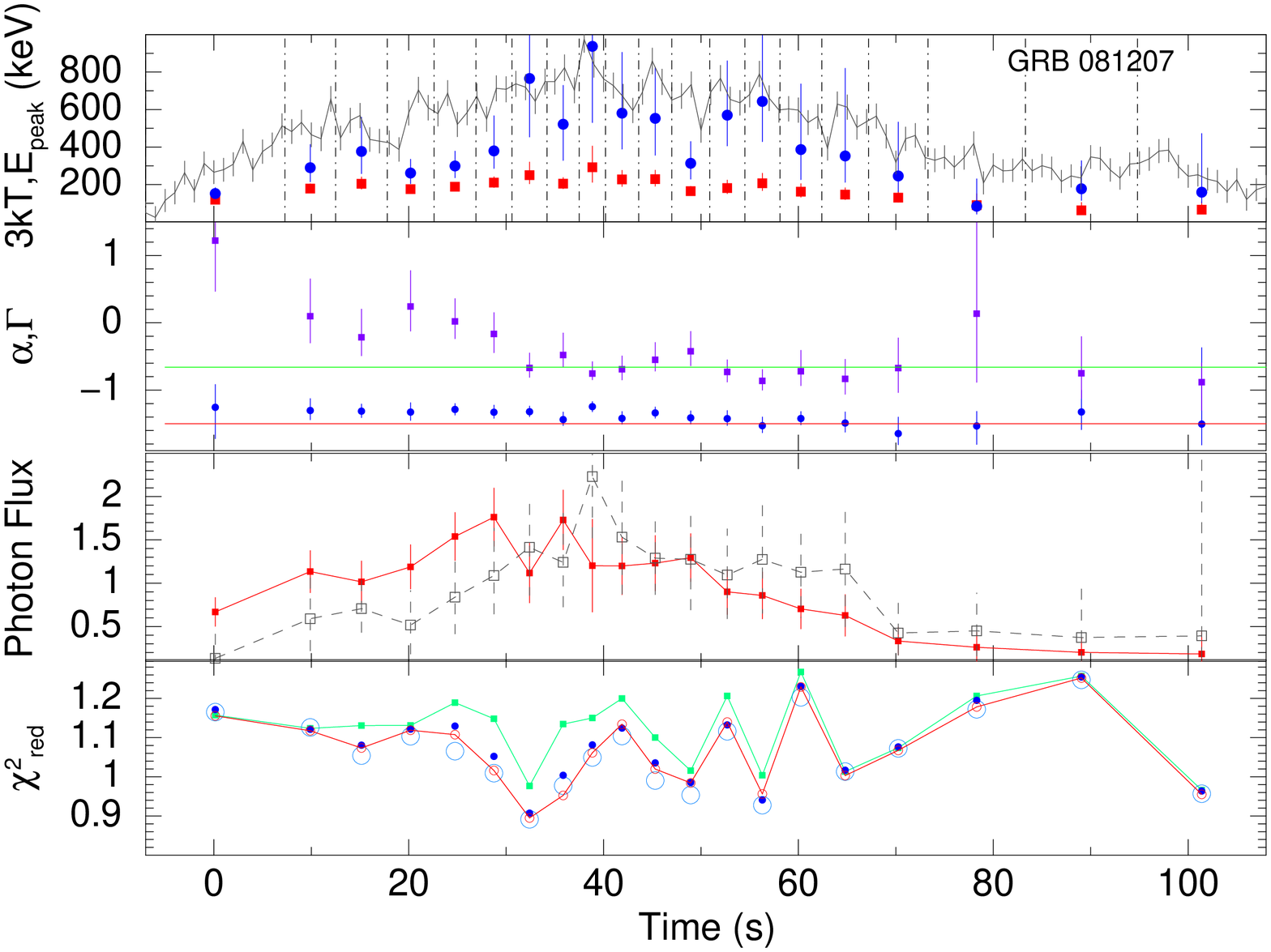} 

}
\caption{GRB 081207. Symbols used are the same as in Figure~\ref{grb1}
}
\label{grb7}
\end{figure*}

\begin{figure*}\centering
{

\includegraphics[width=6.0in]{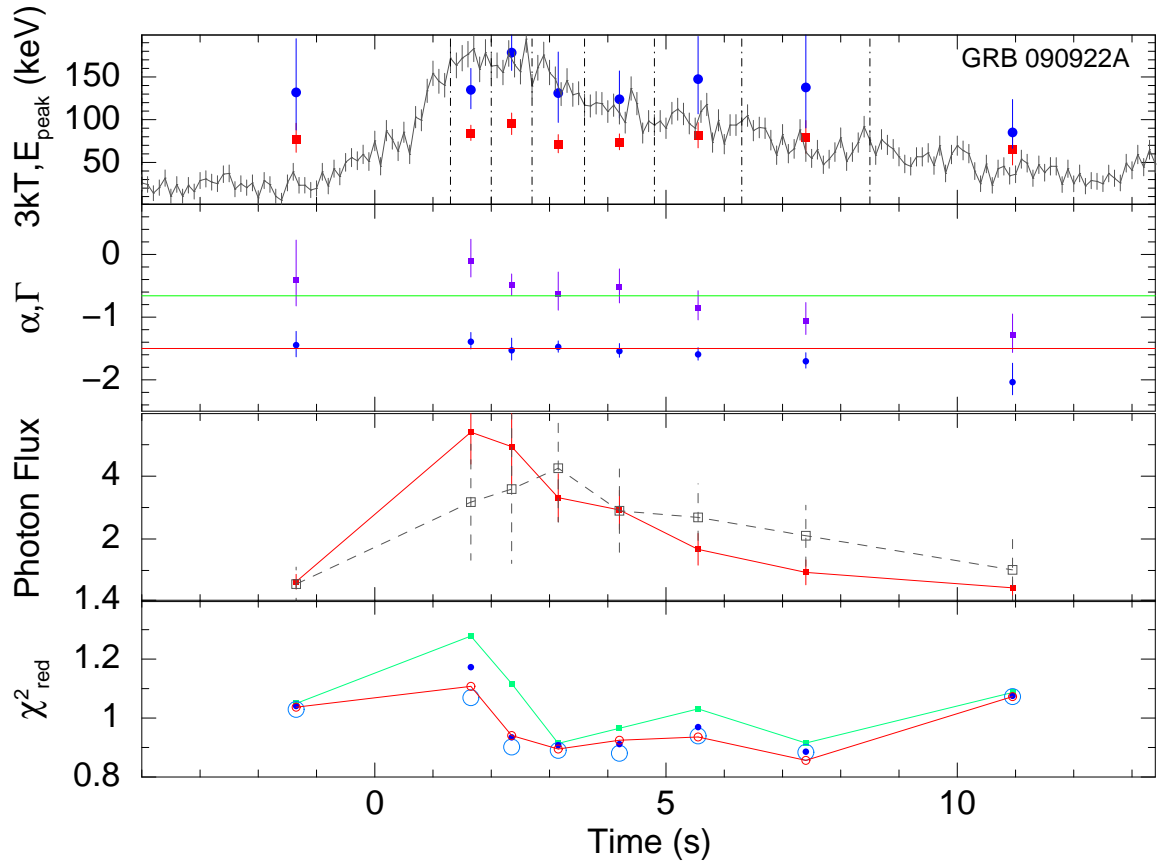} 

}
\caption{GRB 090922A. Symbols used are the same as in Figure~\ref{grb1}
}
\label{grb8}
\end{figure*}

\begin{figure*}\centering
{

\includegraphics[width=6.0in]{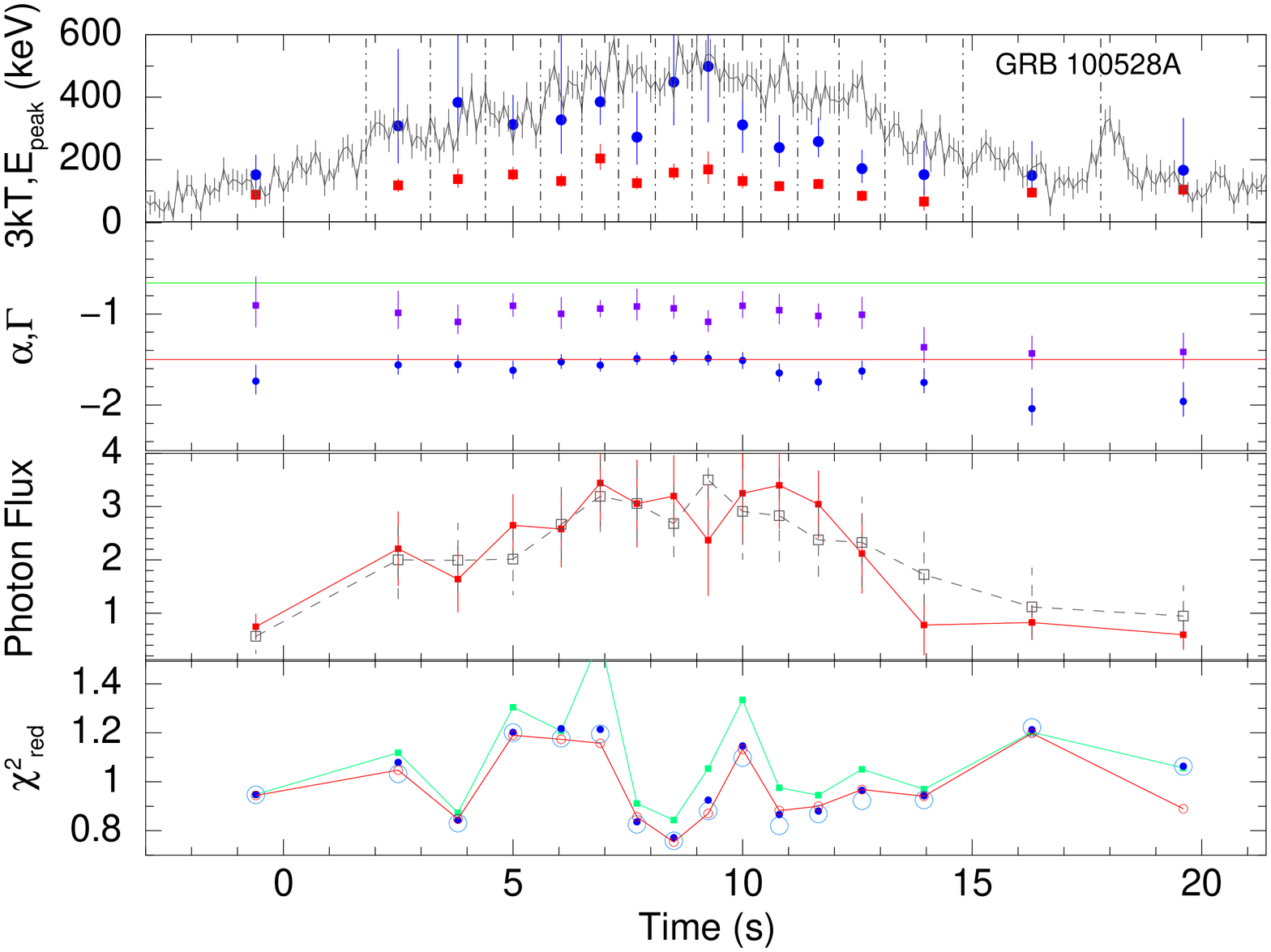} 

}
\caption{GRB 100528A. Symbols used are the same as in Figure~\ref{grb1}
}
\label{grb9}
\end{figure*}

\subsection{Spectral Analysis}
GBM contains 12 sodium iodide (NaI) detectors, numbered as n$x$, where $x$ runs from 0 to 11 (in hexadecimal system, i.e., 
n0 to n9 and then `na' and `nb', where `a' stands for 10 and `b' stands for 11), and 2 bismuth germanate (BGO) 
detectors, denoted similarly as b$y$. The NaI 
detectors cover the lower energy part of the spectrum (8 keV-900 keV), while the BGO covers 200 keV to 40 MeV
(Meegan et al. 2009). The data is supplied in 3 formats: (1) time tag events (TTE) --- time and energy channel information 
as registered by the detector of the individual tagged event (both source and background events) are stored, (2) CSPEC: 
spectrum binned in 1.024 s (during the burst) and 4.090 s (before and after the burst), with
full spectral resolution are stored, and (3) CTIME: binned data in 0.064 s bins
and 8 energy channels are stored. We choose TTE data due to its high time resolution for time-resolved spectroscopy. We examine the
quicklook products from the supplied data and choose three NaI detectors which have registered the highest count rate. 
We choose one of the BGO detectors ---
b0, if $x\le5$ and b1, otherwise. For ambiguous cases, we choose one of the BGO detectors subject to where $x$ occurs 
in most of the cases. For the background estimation, we choose regions before and after the burst, as long as possible, but 
away from the burst. We fit up to fourth order polynomial to model the background, and interpolate the function
to estimate the background during the burst.   

For time-resolved study, we divide the duration of a GRB by requiring a minimum of 1200 background subtracted 
counts per time bin (i.e., 35$\sigma$) for the NaI detector having the highest count rate. We start to integrate the
count from the zero flux level and integrate till this count is reached. The choice of required count 
per time bin is subjective. The choice depends on two competing perspectives: (i) finer time bin may give larger uncertainties in 
the parameters, (ii) wider time bin may not capture the evolution. We made the choice of the time intervals such 
that the variation may be comparable with the uncertainty. We found that this is the case when we use ~0.5 sec 
time bins at the peak. As these GRBs have $\sim2500$ count rate 
at the peak, 1200 count is a good choice. Note that Lu12 also have used the same sigma level.
Only in case of GRB 100707A the requirement is 2500 count/bin (i.e., 50$\sigma$), as the count 
rate of this GRB is much higher than the rest. Note that due to lower effective area of BGO, it is not suitable for time-resolved 
spectroscopy (see e.g., Ghirlanda et al. 2010). The count in the BGO energy range is often less than 2$\sigma$. 
However, it is necessary to use this detector to constrain the high energy part of 
the spectrum. Hence, we use larger bin size for the spectrum in the BGO energy range i.e., 5 or 7 bins are used with progressively 
higher bin size at higher energy. Spectrum of NaI detectors are binned requiring 40 counts per bin. We use {\tt rmfit v3.3pr7}
for lightcurve extraction and background estimation. The time-resolved spectral study is performed using {\tt XSPEC v12.6.0}.
We use $\chi^2$ minimization procedure to estimate the parameters and their nominal 90\% error.
Note that the parameter estimation done by $\chi^2$ technique does not deviate lagrer than 10\% from the c-stat 
technique for a count of 1000 (see Nousek \& Shue 1989). We have used $\chi^2$ technique to compare goodness of fit 
between different models which cannot be done for c-stat method. In Basak \& Rao 2013a, we have conclusively shown 
for a set of GRBs that the deviation of parameter values due to different statistics (i.e., $\chi^2$ and c-stat) is smaller than 
that due to different choice of detectors, background time and modelling etc.

We use four models for time-resolved spectral fitting. These are: (1) Band function, (2) blackbody along with a power-law (BBPL),
(3) a disk blackbody, having continuous temperature profile with radius, along with a power-law (mBBPL) and (4) two blackbodies
along with a power-law (2BBPL). BR13 have used these models for the individual pulses of two GRBs, namely,
GRB~081221 and GRB~090618.

\begin{table*}

\caption{Classification of the GRBs: hard-to-soft (HTS) and intensity tracking (IT) based on the spectral analysis}

\begin{tabular}{cccccccc}
\hline 
GRB & Type & Behaviour of  & Mean $\alpha$ & Deviation of $\alpha$ $^{(a)}$  & $\alpha$ crossing -2/3  & Band/BBPL$^{(b)}$ & LAT detection\\
    &              &   the PL Flux &  & from -3/2 ($\chi^2_{\rm red}$)& ``line of death''  &          &              \\
\hline
\hline
081224	& HTS & Clear delay and lingering & $-0.43 \pm 0.04 $ & 26.7$\sigma$ (53.5) & 11/15 & 0.80 (0.14) & 3.1$\sigma$\\
090809B & HTS & Mild delay and lingering & $-0.64 \pm 0.06 $ & 14.3$\sigma$ (13.5) & 8/15 & 0.74 (0.15) & No\\
100612A & HTS & Very mild lingering & $-0.58 \pm 0.07 $ & 13.1$\sigma$ (19.1) & 7/11 & 0.73 (0.15) & No\\
100707A & HTS & Clear delay and lingering & $0.013 \pm 0.017 $ & 89.0$\sigma$ (71.6) & 16/18 & 0.74 (0.23) & 3.7$\sigma$\\
110721A & HTS & Clear delay and lingering & $-0.95 \pm 0.02 $ & 27.5$\sigma$ (56.2) & 2/15 & 0.94 (0.10) & 30.0$\sigma$\\
110817A & HTS & Mild lingering & $-0.31 \pm 0.07 $ & 16.9$\sigma$ (34.1) & 8/8 & 0.81 (0.09) & No\\
\hline
081207	& IT & Mild delay & $-0.63 \pm 0.06 $ & 14.5$\sigma$ (10.5) & 10/20 & 0.66 (0.12) & No\\
090922A & IT & Mild delay and lingering & $-0.66 \pm 0.10 $ & 8.4$\sigma$ (9.7) & 5/8 & 0.67 (0.13) & No\\
100528A & IT & Mild lingering & $-1.02 \pm 0.04 $ & 12.0$\sigma$ (9.8) & 0/16 & 0.68 (0.14) & No\\
\hline
\end{tabular}
\label{t2}
\begin{flushleft}
\begin{footnotesize}

$^{(a)}$ Deviation of the mean value of $\alpha$ from the -3/2 ``line of death'' (fast cooling) in the units of $\sigma$ is shown. 
$\chi^2_{red}$ is the reduced $\chi^2$ of the fit of $\alpha$ values assuming the model $\alpha=-3/2$. Higher 
$\chi^2_{red}$ denotes higher deviation from the synchrotron model predicted photon index (for fast cooling).
\\
\vspace{0.1in}
$^{(b)}$ Band function compared to the BBPL model. F-test is performed to check the superiority of Band function over 
the BBPL model for all the time-resolved spectrum of each GRB. The reported values are the mean (and standard deviation)
of the confidence level of F-test.
\end{footnotesize}
\end{flushleft}

\end{table*}

\section{Results}
\subsection{Results of Timing Study}
We fit the LCs of all the GRBs using Equation~1. These are shown in Figure~\ref{lc}.
We note that Norris model adequately captures the pulse profile of all the GRBs. However, note that the $\chi^2_{red}$ are
generally not good. This is due to the fact that the pulses, other than the broad structure, 
have very rapid variations (see Rao et al. 2011). These finer time variability cannot be captured 
in a simple pulse structure. In the current study, we are interested to quantify broad pulse properties 
e.g., width and asymmetry of the pulses. Hence, the use of Norris model is adequate for our purpose.  

In Table~\ref{t1}, we report the parameters of Norris model fit. The errors in the derived parameters are 
calculated by propagating the mean error of the Norris model parameters (see Norris et al. 2005). 
Uncertainty in p, however, is obtained by noting down the $\chi^2_{\rm red}$ for different values of p because the
correlated errors in $\tau_1$ and $\tau_2$ gives incorrect results,
particularly for symmetric GRBs.
The width (w) and asymmetry ($\kappa$), however, are well determined. 
Though these GRBs have broad single pulses, we find that for GRB 081207, which is a very long GRB compared 
to all others, at least two heavily overlapping pulses are present during the main bursting episode. Also,
for GRB 110721A, we found that fitting two pulses is significantly better than a single pulse fitting (an 
improvement of $\Delta \chi^2 = 205.1 $ is obtained to the expense of 4 degrees of freedom).

Lu12 have suggested that the HTS pulses are more asymmetric than the IT pulses. But, from Table~\ref{t1} we find that there are
some exceptions to  this consensus for our single pulse sample. For example, GRB~100612A, which is a HTS GRB, is 
very symmetric ($\kappa$ is only $0.14\pm0.02$).
GRB~090922A, despite being an IT GRB, is very asymmetric (as high as $0.62\pm0.05$). The highest asymmetry is shown by 
single pulse GRB~100707A. We visually inspect this GRB and find that the rising part is in fact very steep. Hence, in order
to get more data point in the rising part, we analyze the LC using 0.1 s bin size. We find the asymmetry, $\kappa=0.72\pm0.01$, 
which is indeed the highest among all the GRBs. Though we found two pulses in GRB 110721A, one of these pulses 
shows even higher asymmetry $\kappa=0.83 \pm 0.05$. One of the two pulses of GRB~081207, an IT GRB, show high asymmetry 
($0.57\pm0.06$). Hence, we cannot differentiate between HTS and IT GRBs according to their 
asymmetric LCs.

\subsection{Results of Time-resolved Spectroscopy}
As described in Section 2.2, we extract the time-resolved spectra of the individual GRBs by requiring a 
minimum background subtracted count per time bin. The time interval ($t_1$ to $t_2$) and the number of 
bins (n) are listed in Table~\ref{t1}. For GRB~081207, which has another pulse after 80 s, the data is taken till 76 s. 
This selection, however, will not affect any conclusion regarding this GRB. We also report the detectors used in our analysis 
in the last column of Table~\ref{t1}.

The results of the time-resolved spectroscopy for all the GRBs are shown graphically in Figure~\ref{grb1} through 
Figure~\ref{grb9}. The upper panels show the kT variation of the BBPL and $E_{\rm peak}$ variation of the Band function. 
We have shown the LCs of the individual GRBs in this panel to compare the $E_{\rm peak}$ variation with the 
count rate variation. From this variation we can identify that the first 6 GRBs are the HTS GRBs and the rest 3 are IT GRBs. 
Note that the peak of BB occurs at $\sim$3kT, and hence, in this panel, we have plotted 3kT 
instead of kT to compare the peak position given by the BBPL and Band function fit. We note that there is a general agreement
of evolution of these peaks, however, the peak of the BBPL fit is always found to be lower than the 
corresponding $\nu F_{\nu}$ peak of the Band function.

In the second panels from the top, we have shown the variation of $\alpha$ of the Band function and $\Gamma$ of the BBPL model.
The parameter $\beta$ is often unconstrained or reaches the hard bound of -10.
Hence, we do not show this parameter. In this panel, we show the ``synchrotron lines of death'' by two straight
lines --- one at -2/3 (slow cooling regime) and another at -3/2 (fast cooling regime). The general trend of $\alpha$ 
is higher at the beginning and lower in the later part. Note that the value $\alpha$ is always greater than the line 
at fast cooling regime (-3/2). To quantify the deviation of $\alpha$ from the predicted -3/2 line, we device
two methods: (a) the deviation of the mean value of $\alpha$ from -3/2 in the units of $\sigma$ and (b) the $\chi^2$ value of 
the fit to $\alpha$, assuming the model predicted $\alpha$=-3/2. In Table~\ref{t2} (column 4), we have shown the mean 
values of $\alpha$ for each GRB. The error in the mean is 1$\sigma$, and it is calculated by using the two tailed 
errors of $\alpha$ (at nominal 90\% confidence). The deviation of the mean from -3/2 value in the units of $\sigma$
is shown in column 5. We see a high deviation in each case. The $\chi^2_{\rm red}$ (shown in parenthesis) is also high 
in each case denoting the significance of the trend. In some cases, it's value is greater than -2/3, which is disallowed 
by synchrotron model for slow cooling electrons. This deviation from -2/3 line, however, is not very significant. 
For some HTS GRBs the deviations are significant: 40$\sigma$ (GRB 100707A), 5.8$\sigma$ (GRB~081224) and 5.0$\sigma$ 
(GRB~110817A). For IT GRBs the deviations are consistent with zero (within 1$\sigma$). For IT GRB~100528A, the mean 
value of $\alpha$ is $-1.02\pm0.04$ which is even within the slow cooling ``line of death'' at 8.8$\sigma$.
We find two extreme cases. These are GRB~110817A (HTS) and GRB~100528A (IT). In the former case, $\alpha$ values are 
always above the line at -2/3 and in the latter case $\alpha$ values are always below the line. 

In the third panels, we have shown the evolution of the photon flux for the two components of BBPL model.
Note that, the PL flux lags behind the BB flux. This behaviour is particularly significant for GRB~081224,
GRB~100707A and GRB~110721A. In a recent paper, Basak \& Rao (2013b) have shown that very high energy (GeV) emission is 
expected for GRBs which have delayed PL emission. For the three GRBs in the present study, we also find reported LAT 
detection (LLE data) at 3.1$\sigma$, 3.7$\sigma$ and 30.0$\sigma$, respectively (Ackermann et al. 2013). Though we have
not used the LAT data of these GRBs for spectroscopy, the LAT detection is consistent with our earlier claim.

Finally, the fourth panels show the $\chi^2_{\rm red}$ of various model fits. To compare the fit statistics of BBPL model with
other models, we perform F-test between BBPL and Band function. The confidence level (CL) that the alternative model (Band)
is better than the original hypothesis (BBPL) is calculated for all the time-resolved bins of the individual GRBs.
We then compute the mean and standard deviation of the CL. The corresponding values are reported in Table~\ref{t2}. 
It is evident that the IT pulses are adequately fit by BBPL model. The HTS pulses, on the other hand, require a different model 
to fit the spectrum. However, Band, mBBPL and 2BBPL are all equally good for this purpose. Hence, one needs physical arguments 
to distinguish between these models. 

\begin{figure*}\centering
{

\includegraphics[width=6.5in]{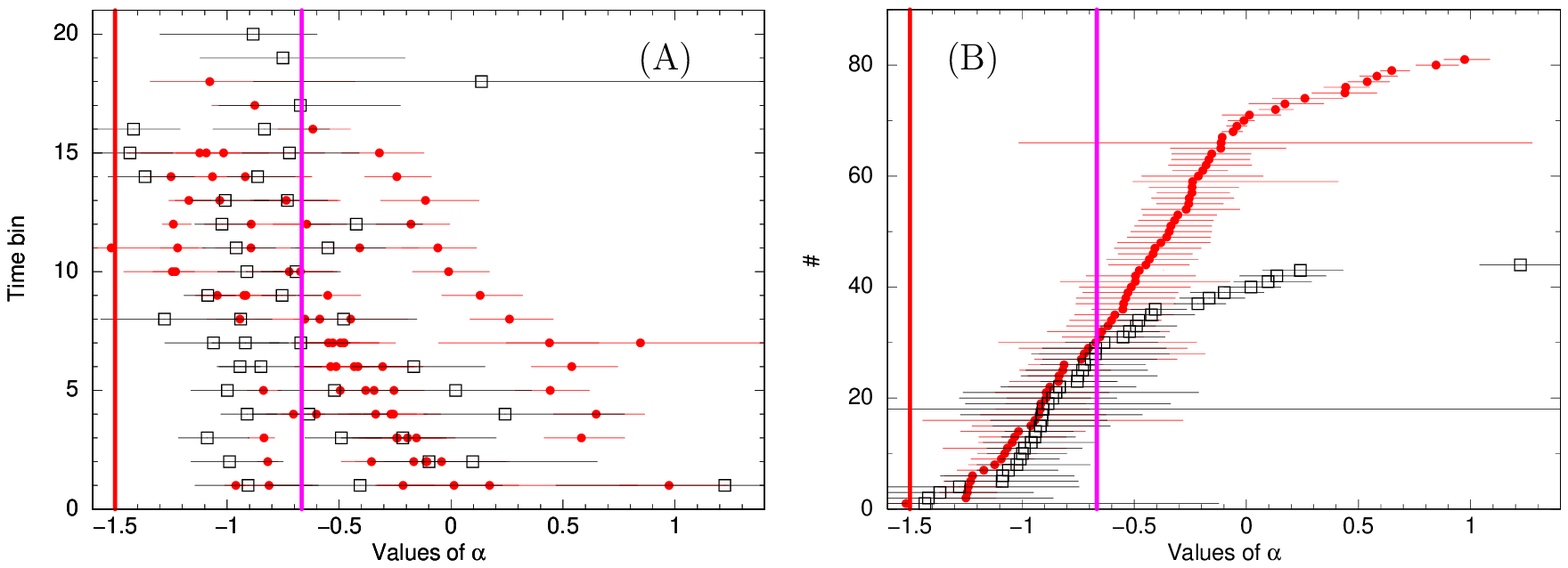} 

}
\caption{(A) The values of $\alpha$ for all GRBs --- filled circles for HTS class and open boxes for the IT class. The errors are
calculated at nominal 90\% confidence. The y-axis denotes the time bin from start to end. The
thick red and purple lines are the lines of death at $\alpha=-3/2$ and $\alpha=-2/3$, 
respectively. (B) Same as A, with values of $\alpha$ sorted in ascending order. 
The values of $\alpha$ are always greater than -3/2 \textbf{(see Table~\ref{t2} for 
significance)}. Note that the region of $\alpha<-2/3$ (i.e., the left
of the purple line) is mainly populated by IT GRBs, while the $\alpha$ values of 
HTS GRBs appear in the region of $\alpha>-2/3$ . The mean value of $\alpha$ are \textbf{-0.42} and -0.68 for HTS and IT
GRBs, respectively
}
\label{alpha_hts_it}
\end{figure*}

The results of all the GRBs are summarized in Table~\ref{t2}. We note that the GRBs can be classified in 2 categories: HTS
and IT. We note the following trends for HTS and IT GRBs. First, note that a LAT detection is reported only for HTS class.
For HTS GRBs, the value of $\alpha$ is greater than -2/3 in 63.4\% cases. The exception of this trend is shown by GRB 110721A.
Excluding this GRB, we obtain 74.6\% of cases of HTS GRBs where $\alpha$ is greater than -2/3. The deviation of $\alpha$
from -3/2 line are systematically high (large $\sigma$ deviation and high values of $\chi^2_{\rm red}$).
The confidence level (CL) that the Band model is better than the BBPL model has a mean value
of 79.3\%. In case of IT GRBs, the value of $\alpha$ is greater than -2/3 in only 44\% cases. 
As pointed out the mean values are greater only within 1$\sigma$ deviation, and for one IT GRB (namely, 100528A)
it is even less than -2/3 at 8.8$\sigma$. Also, the CL that the Band model
is better than BBPL model has a mean value of only 67\%. The only exception of the behaviour of $\alpha$ is noted for
the HTS GRB 090809B for which $\alpha$ is greater than -2/3 in only 53.3\% cases. However, note in Figure~\ref{grb2} (upper
panel) that the $\rm E_{peak}$ evolution is not strictly hard-to-soft.
In Figure~\ref{alpha_hts_it} (left panel), we have plotted the $\alpha$ values for HTS and IT GRBs, with y-axis as time sequence.
In the same plot, we have indicated the -3/2 and -2/3 lines of death. Note that $\alpha$ values are always greater than
the -3/2 line. In many cases the $\alpha$ values are greater than the -2/3 line, especially for the HTS GRBs. The mean value
of $\alpha$ is -0.42 and -0.68 for HTS and IT GRBs with dispersion of 0.72 and 0.50, respectively.
Also there is a clear trend of the evolution of $\alpha$ --- the values of $\alpha$ decreases with time i.e., the
spectrum becomes softer.
\section{Discussion}

\subsection{Features of 2BBPL model}
\begin{figure*}\centering
{

\includegraphics[width=6.0in]{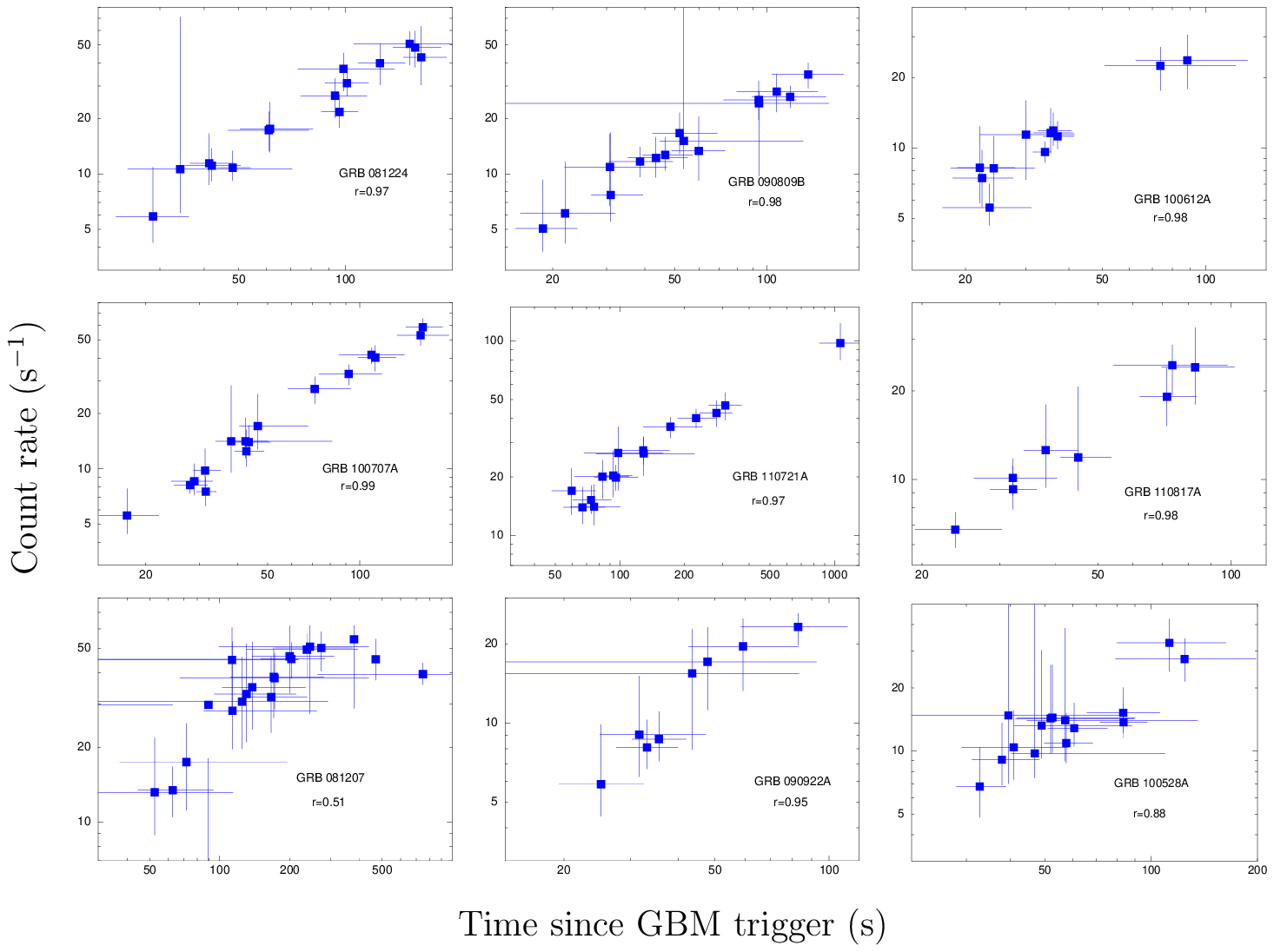} 

}
\caption{Correlation between the temperatures of the two BBs of 2BBPL model. The upper 6 panels are for HTS GRBs and 
the other 3 panels are for IT GRBs. We note significant correlation in all the cases.
}
\label{kt_corr}
\end{figure*}

The four models used in our analysis are taken from BR13, who applied them to the brightest GRBs having separable pulses 
(GRB~081221 and GRB~090618). The present results generally agree with the results 
obtained for those two bright GRBs. In the present case too we find that the decomposition
of the spectrum in terms of 2BBs or mBB and a power-law components significantly improves the
fit over single BB with a PL. Though the 2BBPL model gives better $\chi^2_{red}$ in all cases, the 
present analysis cannot conclusively show whether the 2BBPL model is an approximation
for some underlying complex spectral distribution. Exploring this model 
further,  Basak \& Rao (2013b), have  shown some very promising results like 
prediction of LAT emission from GBM only analysis. They have fitted the MeV GBM data, without invoking the LAT data, with the 
2BBPL model and have shown that the PL flux, independent of the LAT data, mimics the LAT GeV emission. The PL has a delayed
onset and lingers at the later part, just as the same way as the LAT flux. Moreover, the PL fluence correlates with the LAT 
fluence. In the present analysis, we find LAT detection for three GRBs (GRB~081224, GRB~100707A and GRB~110721A) which 
show similar behaviour (a delayed onset of the power-law component which lingers at the later phase). 
Another interesting feature  we find that the 2BBs of the 2BBPL model are highly correlated. In Figure~\ref{kt_corr}, we have plotted the 
temperatures of the two BBs of the 2BBPL model. It is clear that the two temperatures are highly correlated. This correlation shows that either there is some underlying 
physical reason for the two black body components or they are connected to each other
because they happen to be an approximation for some more complex spectral shape.

\section{Summary}
To summarize, we have collected a sample of GBM GRBs having single pulse. We have studied the individual pulses in the time 
and energy domain and found some interesting results. We found that the Band model always gives better fit to the data compared
to the BBPL model. The $\chi^2_{red}$ of the Band model is comparable to mBBPL and 2BBPL model. However, the values of
$\alpha$, in many cases, are greater than -2/3, the limit due to the synchrotron emission from electrons in the slow cooling regime.
Hence, the spectrum cannot be fully synchrotron. Hence, neither BBPL nor Band can be considered consistent with the underlying physical 
mechanism of the prompt emission. It is possible that both thermal and non-thermal emissions contribute 
to the total emission. In this regard, the 2BBPL and the mBBPL model are preferred over BBPL model.
We found that the HTS GRBs have generally larger values of $\alpha$ compared to the IT GRBs. 
Also, the $\alpha$ has a decreasing trend with time, making the spectrum softer. This signifies that the spectrum
has a different origin than synchrotron at the beginning, but later synchrotron emission may dominate. 

We found that the peak energy of the BBPL model gives lower value than the corresponding Band function. 
As the peak energy of the spectrum is used for the correlation studies in the prompt emission (Amati et 
al. 2002; Yonetoku et al. 2004; Ghirlanda et al. 2004; Basak \& Rao et al. 2012a), 
it is essential to determine the correct peak energy of the spectrum. If the underlying reason for the  $E_{\rm peak}$ and 
isotropic equivalent energy correlation ($E_{\rm peak}$-$E_{\gamma, \rm iso}$: the Amati correlation) 
is some basic physical process, a correct model description 
and a proper identification of $E_{\rm peak}$ would be required to improve the Amati-type correlation 
(note that $E_{\rm peak}$ and 3 KT are note strictly correlated with each other and the difference varies between GRBs). Investigating
a pulse-wise Amati correlation (e.g., see Krimm et al. 2009; Basak \& Rao 2013c) with different models to 
derive the $E_{\rm peak}$ can provide inputs to strengthen the Amati correlation 
as well as to identify the physical process responsible for the correlation. Note that 
a HTS evolution, by its nature violate time-resolved $E_{\rm peak}$-$E_{\gamma, \rm iso}$
correlation, because the high values of $E_{\rm peak}$ is obtained even when the flux 
(and hence $E_{\gamma, \rm iso}$) is low. For example, Axelsson et al. (2012) have found $E_{\rm peak}\sim 15$ MeV
during the initial bins of GRB 110721A. Hence, time-resolved Amati correlation is likely to fail. Basak \& Rao (2012b)
have studied Amati correlation by replacing time-resolved $E_{\rm peak}$ and $E_{\gamma, \rm iso}$ by pulse average $E_{\rm peak}$ and 
total $E_{\gamma, \rm iso}$ of a pulse. A pulse average $E_{\rm peak}$, due to the crucial averaging over a pulse,
will not show the effect of HTS evolution history, thus restoring the Amati correlation. Note that, pulses, rather than 
time-resolved bins are preferred as the pulses behave as independent entities in GRBs. In this context 
one reasonable choice is peak energy at the zero fluence ($E_{\rm peak,0}$) as done by Basak \& Rao (2012b). Unlike 
$E_{\rm peak}$, which is a pulse average quantity, $E_{\rm peak,0}$ is a constant for each pulse and correlates better with the total 
$E_{\gamma, \rm iso}$. Consequently, the strong pulse-wise $E_{\rm peak,0}$-$E_{\gamma, \rm iso}$ correlation indicates 
that a pulse with high initial peak energy may have started with a low flux, but eventually it will give rise to high total flux
of the pulse. However, one cannot use the IT class for such correlation study, as 
IT GRBs, by their nature will not give such $E_{\rm peak,0}$. Hence, one can use pulse average peak energy 
for such correlation study, which does not depend on HTS or IT evolution (Basak \& Rao 2013c)

Finally, we want to remind that the 2BBPL model has some interesting features like highly 
correlated BB temperatures. The PL component of this model fitted to the prompt keV-MeV emission has been shown 
to correlate with the high energy GeV emission (Basak \& Rao 2013b). These definitely point towards some 
underlying physical process. Hence, it is important to find the existence of the 2BBs in the prompt 
emission sepctrum. Making similar spectral analysis for GRBs with better low energy measurement (like that 
from \textit{Swift}/X-Ray Telescope (XRT))  would clarify these further. As the temperature of the two correlated BBs 
decrease with time, the lower BB may show up in the low energy detector like XRT during the late prompt emission phase.
However, such detailed study needs good knowledge of detector systematics and cross-detector calibration. In Basak \& Rao (2012b),
we have shown that spectral analysis using BAT and GBM gives systematics difference in the results. Though one needs to
find out ways to perform joint fitting, the procedure is quite involved and beyond the scope of the present paper.

\section*{Acknowledgments} This research has made use of data obtained through the
HEASARC Online Service, provided by the NASA/GSFC, in support of NASA High Energy
Astrophysics Programs. We thank the referee for valuable comments and numerous suggestions
especially to make quantitative statements of the results.

\end{document}